\documentclass[prl,twocolumn,showpacs]{revtex4-1}
\usepackage{amsmath}
\usepackage{amssymb}
\usepackage{graphicx}

\newcommand{\be}[1]{\begin{equation}\label{#1}}
\newcommand{\ee}{\end{equation}}
\newcommand{\ba}[1]{\begin{eqnarray}\label{#1}}
\newcommand{\ea}{\end{eqnarray}}

\begin{document}

\title{Linking dissipation-induced instabilities with nonmodal
growth: the case of helical magnetorotational instability}

\author{G. Mamatsashvili$^{1,2,3}$}
\email{g.mamatsashvili@hzdr.de}
\author{F. Stefani$^1$}
\email{f.stefani@hzdr.de} \affiliation{$^1$Helmholtz-Zentrum
Dresden-Rossendorf, P.O. Box 510119, D-01314 Dresden, Germany\\
$^2$Department of Physics, Faculty of Exact and Natural Sciences, Tbilisi State University, Tbilisi 0179, Georgia\\
$^3$Abastumani Astrophysical Observatory, Ilia State University,
Tbilisi 0162, Georgia}

\date{\today}

\begin{abstract}
The helical magnetorotational instability is known to work for
resistive rotational flows with comparably steep negative or
extremely steep positive shear. The corresponding lower and upper
Liu limits of the shear are continuously connected when some axial
electrical current is allowed to flow through the rotating fluid.
Using a local approximation we demonstrate that the
magnetohydrodynamic behavior of this dissipation-induced instability
is intimately connected with the nonmodal growth and the
pseudospectrum of the underlying purely hydrodynamic problem.
\end{abstract}

\pacs{47.32.-y, 47.35.Tv, 47.85.L-, 97.10.Gz, 95.30.Qd}

\maketitle

The magnetorotational instability (MRI) \cite{VELI} is believed to
trigger turbulence and enable outward transport of angular momentum
in magnetized accretion disks \cite{BAHA}. The typical Keplerian
rotation of the disks belongs to a wider class of flows with
decreasing angular velocity and increasing angular momentum, which
are Rayleigh-stable \cite{RAYLEIGH}, but susceptible to the standard
version of MRI (SMRI), with a vertical magnetic field $B_z$ imposed
on the rotating flow. For SMRI to operate, both the rotation period
and the Alfv{\'{e}}n crossing time have to be shorter than the
timescale for magnetic diffusion  \cite{LIU2006}. For a disk of
scale height $H$, this implies that both the magnetic Reynolds
number ${\rm Rm}=\mu_0 \sigma H^2 \Omega$ and the Lundquist number
$S=\mu_0 \sigma H v_{A}$ must be larger than one ($\Omega$ is the
angular velocity, $\mu_0$ the magnetic permeability, $\sigma$ the
conductivity, $v_A$ the Alfv{\'{e}}n velocity).

These conditions are safely fulfilled in well-conducting parts of
accretion disks. However, the situation is less clear in  the ``dead
zones'' of protoplanetary disks, in stellar interiors, and in the
liquid cores of planets, because of low magnetic Prandtl numbers
${\rm Pm}=\nu/\eta$ there \cite{BH08}, i.e. the ratio of viscosity
$\nu$ to magnetic diffusivity $\eta=(\mu_0 \sigma)^{-1}$. Moreover,
in compact objects like stars and planets even the condition of
decreasing angular velocity is not everywhere fulfilled: an
important counter-example is the equator-near strip (approximately
between $\pm 30^{\circ}$) of the solar tachocline \cite{PAME}, which
is, interestingly, also the region of sunspot activity \cite{CHA}.

The helical version of MRI (HMRI) is interesting both with respect
to the low-${\rm Pm}$ problem as well as for regions with positive
shear. Adding an azimuthal magnetic field $B_{\phi}$ to $B_{z}$,
Hollerbach and R\"udiger \cite{HR95} had shown that this
dissipation-induced instability works also in the inductionless
limit, ${\rm Pm}=0$, and scales with the Reynolds number ${\rm
Re}={\rm Rm} {\rm Pm}^{-1}$ and the Hartmann number ${\rm Ha}=S {\rm
Pm}^{-1/2}$, in contrast to SMRI that is governed by ${\rm Rm}$ and
$S$. Soon after, Liu et al. \cite{LIU} showed that HMRI is
restricted to rotational flows with negative shear slightly steeper
than the Keplerian, or extremely steep positive shear. Specifically,
their short-wavelength analysis gave a threshold of the negative
steepness of the rotation profile $\Omega(r)$, expressed by the
Rossby number ${\rm Ro}= r(2\Omega)^{-1} \partial \Omega/
\partial r$, of ${\rm Ro}_{\rm LLL}=2(1{-}\sqrt 2)\approx -0.828$,
and a corresponding threshold of the positive shear, at ${\rm
Ro}_{\rm ULL}=2(1{+}\sqrt 2)\approx 4.828$. Here, the abbreviations
LLL and ULL refer to the lower and upper Liu limits, respectively.

Surprisingly, the same Liu limits were later found
\cite{APJ12,KS1314} to apply also to the so-called azimuthal MRI
(AMRI) -- a non-axisymmetric ''sibling'' of the axisymmetric HMRI
that prevails for large ratios of $B_{\phi}$ to $B_{z}$
\cite{TEELUCK}. Quite recently, the destabilization of steep
positive shear profiles by purely azimuthal fields was demonstrated
by means of both a short-wavelength analysis \cite{SK15} and a
one-dimensional stability analysis for a Taylor-Couette flow with
narrow gap \cite{RUEDIGER}.

By allowing axial electrical currents not only at the axis, but also
within the fluid, i.e. by enabling the radial profile $B_{\phi}(r)$
to deviate from the current-free case $\propto 1/r$, it was recently
shown \cite{KS1314} that the LLL and the ULL are just the endpoints
of one common instability curve in a plane that is spanned by ${\rm
Ro}$ and a corresponding steepness of the azimuthal magnetic field,
called magnetic Rossby number, ${\rm Rb}=r (2 B_{\phi}/r)^{-1}
\partial{(B_{\phi}/r)}/ \partial r$. In the limit of large ${\rm
Re}$ and ${\rm Ha}$, this curve acquires the closed and simple form
\begin{equation}
{\rm Rb}=-\frac{1}{8}\frac{({\rm Ro}+2)^2}{{\rm Ro}+1}.
\label{rel}
\end{equation}
An interesting consequence of this curve is that the strictness of
the lower Liu limit ${\rm Ro}_{\rm LLL}=-0.828$, which would prevent
Keplerian profiles ${\rm Ro}_{\rm Kep}=-0.75$ from being
destabilized by HMRI or AMRI, could be relaxed if only a small
amount of the axial current is allowed to pass through the liquid.
This effect is now to be investigated in a planned liquid sodium
Taylor-Couette experiment \cite{DRESDYN}, which will combine and
enhance the previous experiments on HMRI \cite{PRE}, AMRI
\cite{SEIL2} and the kink-type Tayler-instability \cite{SEIL1}.

Apart from these interesting theoretical and experimental
achievements, the very existence of the two Liu limits (and the
shape of their connecting curve Eq. (1) in the ${\rm Ro}-{\rm Rb}$
plane) has remained an unexplained conundrum. This Letter aims at
explaining these magnetohydrodynamic features by analysing the
dynamics of HMRI from the nonmodal point of view, which has not been
done before, and linking them to the nonmodal dynamics of
perturbations in the purely hydrodynamic case.

The nonmodal approach to the stability analysis of shear flows in
its most general formulation focuses on the finite-time dynamics of
perturbations, accounting for transient phenomena due to the
shear-induced nonnormality of the flow
\cite{Trefethen92,Farrell1996,Schmid_Henningson2001,Schmid2007}, in
contrast to the canonical modal approach (spectral expansion in
time), which is concerned with behavior at asymptotic times. It
consists in calculating the optimal initial perturbations with a
given positive norm that lead to the maximum possible linear
amplification during some finite time. In self-adjoint flow
problems, the perturbations that undergo the largest amplification
are essentially the most unstable normal modes. By contrast, the
situation is nontrivial in non-selfadjoint shear flow problems: the
normal mode eigenfunctions are nonorthogonal due to the
nonnormality, resulting in transient, or nonmodal growth of
perturbations, which can be substantially faster than that of the
most unstable normal mode \cite{Schmid_Henningson2001,Squire2014}.
So, leaving the effects of the nonnormality out of consideration and
relying only on the results of modal analysis leads to an incomplete
picture of the overall dynamics (stability) of shear flows.

Our main goal is to examine the nonmodal dynamics of HMRI in
differentially rotating flows, which represent a special class of
shear flows for which the nonnormality inevitably plays a role. This
can result in growth factors over intermediate (dynamical/orbital)
times large compared to the modal growth of HMRI. Recently, the
nonmodal dynamics of SMRI was studied by Squire \& Bhattacharjee
\cite{Squire2014} and Mamatsashvili et al. \cite{Mamatsashvili2013};
the present study extends these investigations to the highly
resistive, or low-${\rm Pm}$ regime, where only HMRI survives.

We start with the basic equations of nonideal magnetohydrodynamics
for incompressible conductive media,
\begin{equation}
\frac{\partial {\bf u}}{\partial t}+{\bf u}\cdot \nabla {\bf
u}=-\frac{1}{\rho}\nabla \left(p+\frac{{\bf B}^2}{2\mu_0}\right)
+\frac{{\bf B}\cdot\nabla {\bf B}}{\mu_0 \rho} + \nu\nabla^2 {\bf
u},
\end{equation}
\begin{equation}
\frac{\partial {\bf B}}{\partial t}=\nabla\times \left( {\bf
u}\times {\bf B}\right)+\eta\nabla^2{\bf B},
\end{equation}
\begin{equation}
\nabla\cdot {\bf u}=0,~~~\nabla\cdot {\bf B}=0.
\end{equation}
where $\rho$ is the constant density, $p$ is the thermal pressure,
${\bf u}$ is the velocity and ${\bf B}$ is the magnetic field.

An equilibrium flow represents a fluid rotating with angular
velocity $\Omega(r)$ and threaded by a magnetic field, which
comprises a constant axial component $B_{0z}$ and an azimuthal one
$B_{0\phi}$ with an arbitrary radial dependence:
\[
{\bf u}_0=r\Omega(r){\bf e}_{\phi}, ~~~~{\bf B}_0=B_{0\phi}(r){\bf
e}_{\phi}+B_{0z}{\bf e}_z.
\]
Consider now small axisymmetric ($\partial /\partial \phi=0$)
perturbations about the equilibrium, ${\bf u}'={\bf u}-{\bf u}_0$,
$p'=p-p_0$, ${\bf B}'={\bf B}-{\bf B}_0$. Following
\cite{Pessah2005,LIU,KS1314} we adopt a local (WKB) approximation in
the radial direction around some fiducial radius $r$, i.e., assume
perturbation lengthscales much shorter than the characteristic
lengths of radial variations of the equilibrium quantities, and
represent perturbations as ${\bf u}',{\bf B}'\propto \exp({\rm
i}k_rr+{\rm i}k_zz)$, with axial $k_z$ and large radial $k_r$
wavenumbers, $rk_r\gg1$. Linearizing Eqs. (2)-(4) about the
equilibrium and normalizing time by $\Omega^{-1}$, we arrive at the
following equations for the perturbations (primes are omitted and
the factor $(\mu_0\rho)^{-1/2}$ is absorbed in the magnetic field)
${\boldsymbol \psi}\equiv (u_r, u_{\phi}, B_r, B_{\phi}$) (see
\cite{KS1314,Squire2014} for details):
\begin{equation}
\frac{d\boldsymbol \psi}{dt}={\bf A}\cdot {\boldsymbol \psi},
\end{equation}
where the evolution matrix operator ${\bf A}$ is independent of time
for axisymmetric perturbations and reads as
\[
{\bf A}=\begin{pmatrix}
-\frac{1}{\rm Re} & 2\alpha^2 & {\rm i}\omega_z &  -2\omega_{\phi}\alpha^2 \\
-2(1+{\rm Ro}) & -\frac{1}{\rm Re} & 2\omega_{\phi}(1+{\rm Rb}) &
{\rm
i}\omega_z \\
 {\rm i}\omega_z & 0 & -\frac{1}{\rm Rm} & 0 & \\
-2\omega_{\phi}{\rm Rb} & {\rm i}\omega_z & 2{\rm Ro} &
-\frac{1}{\rm Rm},
\end{pmatrix}
\]
where $\alpha=k_z/k$, $k^2=k_r^2+k_z^2$, $\omega_z\equiv
k_zB_{0z}/\Omega$ and $\omega_{\phi}\equiv B_{0\phi}/r\Omega$. The
Reynolds number, ${\rm Re}=\Omega/\nu k^2$, and the magnetic
Reynolds number, ${\rm Rm}=\Omega/\eta k^2$ are chosen as ${\rm
Re}=4000$ and ${\rm Rm}=0.012$, to give a small magnetic Prandlt
number ${\rm Pm}={\rm Rm}/{\rm Re}=3\cdot 10^{-6}$ typical for
liquid metals and also protoplanetary disks \cite{BH08}. The
strength of the imposed axial field is measured by the Hartmann
number ${\rm Ha}=\omega_z\sqrt{{\rm Re}\cdot {\rm Rm}}$, which is
fixed to ${\rm Ha}=15$ as typical for liquid metal experiments
\cite{PRE,SEIL2}, and the azimuthal field by
$\beta=\omega_{\phi}/\omega_z$. HMRI is most effective in the
presence of an appreciable azimuthal field together with the axial
one, $\beta \sim 1$ \cite{HR95,LIU, KS1314}. We consider
Rayleigh-stable rotation with ${\rm Ro}> -1$ and ${\rm Rb}<0$, since
the axial current decreases with radius. It is readily shown that
${\bf A}$ is indeed nonnormal, or non-selfadjoint, i.e., ${\bf
A}^{\dag}\cdot {\bf A}-{\bf A}\cdot {\bf A}^{\dag}\neq 0$ and the
degree of the nonnormality increases for higher shear ($|{\rm
Ro}|$).
\begin{figure}
\includegraphics[width=\columnwidth, height=6.0cm]{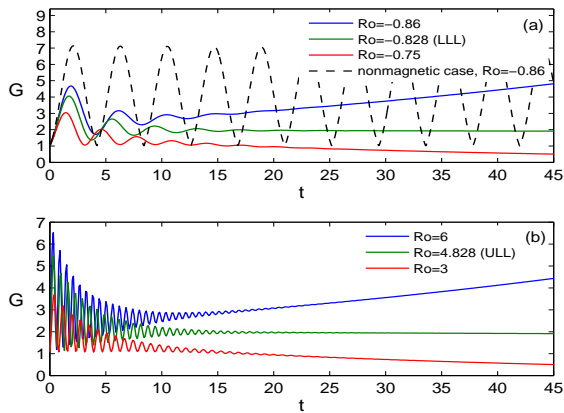}
\caption{Maximum growth $G(t)$ vs. $t$ at different (a) ${\rm
Ro}=-0.86, -0.828 ({\rm LLL}), -0.75 ({\rm Kepler})$ and (b) ${\rm
Ro}=3, 4.828 ({\rm ULL}), 6$. Other parameters are $\alpha=1,
Rb=-1$.}
\end{figure}
\begin{figure}
\includegraphics[width=\columnwidth, height=6.0cm]{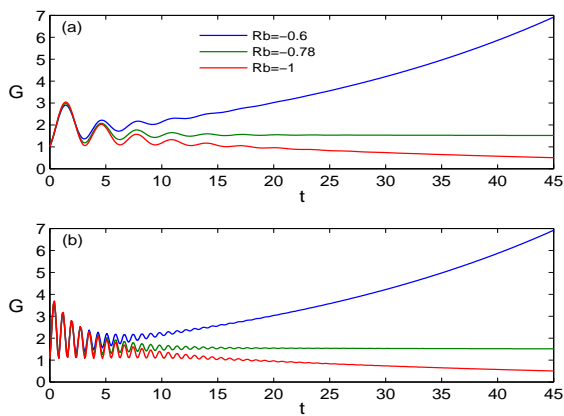}
\caption{$G$ vs. $t$ at different ${\rm Rb}=-1, -0.78, -0.6$ and at
fixed (a) ${\rm Ro}_{\rm Kep}=-0.75$ and (b) ${\rm Ro}=3$ both with
$\alpha=1$.}
\end{figure}

We quantify the nonmodal amplification in terms of the total
perturbation energy, $E=\frac{\rho}{2}(|{\bf u}|^2+|{\bf
B}|^2)={\boldsymbol \psi}^{\dag}\cdot F^{\dag}F\cdot{\boldsymbol
\psi}$, where ${\bf F}=\sqrt{\rho/2}\cdot
diag(\alpha^{-1},1,\alpha^{-1},1)$, which is a physically relevant
norm. The maximum possible, or optimal growth at a specific time $t$
is defined as the ratio $G(t)=\max_{\boldsymbol \psi(0)} E(t)/E(0)$,
where $E(t)$ is the energy at $t$ and the maximization is done over
all initial states ${\boldsymbol \psi}(0)$ with a given initial
energy $E(0)$ (e.g., Ref. \cite{Schmid_Henningson2001}). The final
state at $t$ is found from the initial state at $t=0$ by solving
linear Eq. (5) and can be formally written as ${\boldsymbol
\psi}(t)={\bf K}(t)\cdot {\boldsymbol \psi}(0)$, where ${\bf K}(t)$
is the propagator matrix. Then, the maximum possible amplification
$G(t)$ is usually calculated by means of the singular value
decomposition technique of ${\bf K}$ (e.g., Refs.
\cite{Farrell1996,Schmid_Henningson2001,Trefethen2005,Schmid2007}),
which we adopt here. The square of the largest singular value gives
the value of $G(t)$ and the corresponding initial condition that
achieves this growth (i.e., optimal perturbations) at $t$ is given
by the right singular vector of ${\bf K}$. Finally, we would like to
stress that studying shear flow stability using the nonmodal
approach combined with the method of optimal perturbations is the
most general way of analyzing their dynamics at all times, as
opposed to the modal approach, which concentrates only on the
asymptotic behavior at large times and hence omits an important
class of finite-time transient phenomena.
\begin{figure}
\includegraphics[width=\columnwidth, height=6.5cm]{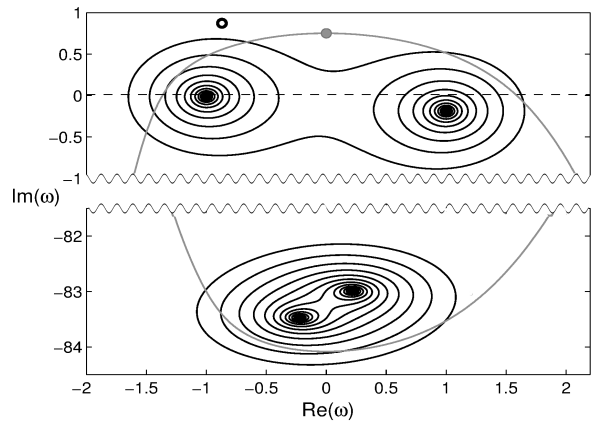}
\caption{Isolines at $\epsilon=10^{0.25}, 10^{0.4},
10^{0.55},...,10^{3.1}$ show the $\epsilon$-pseudospectra of the
${\bf A}$ matrix in the complex $\omega$-plane for ${\rm Rb}=-1,
\alpha=1$ and ${\rm Ro}_{\rm Kep}=-0.75$. The circle indicates the
complex $\omega_K$ corresponding to the Kreiss constant. The gray
curve shows numerical range and the dot on it is the numerical
abscissa. Four black dots represent the eigenvalues of the normal
modes.}
\end{figure}

The modal analysis in the WKB approximation yields an expression for
the growth rate of HMRI in the relevant limit of small ${\rm Pm}$,
but both large ${\rm Re}$ and $\rm Ha$ \cite{LIU, KS1314}. When
maximized with respect to $\beta$ (which is typically around unity),
this growth rate, given by Eq. (8.30) of \cite{KS1314}, becomes (in
units of $\Omega$)
\begin{equation}
\gamma=-\frac{{\rm Ha}^2}{\rm Re}\left[\frac{({\rm
Ro}+2)^2}{8(1+{\rm Ro}){\rm Rb}}+1\right],
\end{equation}
while the real part of the eigenfrequency is equal to the frequency
of inertial waves, $\omega_{\rm iw}=2\alpha\sqrt{1+{\rm Ro}}$.
Equation (6) yields the stability boundary Eq. (1) which indicates
that for ${\rm Rb}=-1$ the instability (i.e., $\gamma>0$) exists at
negative, ${\rm Ro} < {\rm Ro}_{\rm LLL}=-0.828$, and positive,
${\rm Ro} > {\rm Ro}_{\rm ULL}=4.828$, shear, while at larger
$-1<{\rm Rb}<0$, the stability region shrinks and the instability
extends beyond the Liu limits. As a result, the modal growth of HMRI
can also exist for the Keplerian rotation (${\rm Ro}_{\rm
Kep.}=-0.75$) starting from ${\rm Rb}=-0.781$ \cite{KS1314}.

Now we examine the nonmodal growth of HMRI as a function of time.
Figures 1 and 2 show the maximum energy growth $G(t)$ at modally
stable and unstable Rossby and magnetic Rossby numbers together with
the growth in the modally stable nonmagnetic case, where only the
nonmodal growth is possible. In all cases, the initial stage of
evolution is qualitatively similar: the energy increases with time,
reaches a maximum $G_m$ and then decreases. This first nonmodal
amplification phase is followed by minor amplifications. Like in the
case of modal growth, the kinetic energy dominates over the magnetic
one also during nonmodal growth. As a result, the duration of each
amplification event is set by inertial waves and is about the half
of their period. Correspondingly, the peak value $G_m$ is attained
at around one quarter of the period, $t_m\approx \pi/2\omega_{\rm
iw}$, similar to that in the nonmagnetic case, although its value is
smaller than that in the latter case. At larger times, the optimal
growth follows the behavior of the modal solution -- it increases
(for ${\rm Ro}=-0.86, 6$), stays constant (for the Liu limits, ${\rm
Ro}={\rm Ro}_{\rm LLL}, {\rm Ro}_{\rm ULL}$) or decays (for ${\rm
Ro}=-0.75,3$), respectively, if the flow is modally unstable,
neutral or stable; in the latter case HMRI undergoes only transient
amplification. This is readily understood: at large times the least
stable modal solution (with growth rate given by Eq. 6) dominates,
whereas at small and intermediate times the transient growth due the
interference of nonorthogonal eigenfunctions is important. In
particular, for the Liu limits, where the modal growth is absent,
there is moderate nonmodal growth $G_m({\rm Ro_{LLL}})=4.06,
G_m({\rm
Ro_{ULL}})=5.46$. 
A similar evolution of axisymmetric perturbations' energy with time
for HMRI was already found in \cite{Priede2007}, where also the
physical mechanism of HMRI was explained in terms of an additional
coupling between meridional and azimuthal flow perturbations.
Importantly, in Fig. 1, $G_m$ at modally stable and unstable Rossby
numbers are comparable and several times larger than the modal
growth factors during the same time $t_m$. Indeed, at ${\rm
Ro}=-0.86$ the growth achieves the first peak $G_m=4.68$ at
$t_m=1.86$, while at this time the energy of the normal mode would
have grown only by a factor of $\exp[2t_m\gamma({\rm Ro})]=1.034$.
This also implies that in the Keplerian regime, where there is no
modal growth of HMRI for ${\rm Rb} = -1$, it still exhibits moderate
nonmodal growth (red curves in Figs. 1a and 2a). It is seen from
Fig. 2 that the peak $G_m$ is almost insensitive to ${\rm Rb}$,
however, its effect becomes noticeable as time passes. Decreasing
the slope at a given ${\rm Ro}$ increases the optimal growth and at
large times makes the flow modally unstable.
\begin{figure}
\includegraphics[width=\columnwidth]{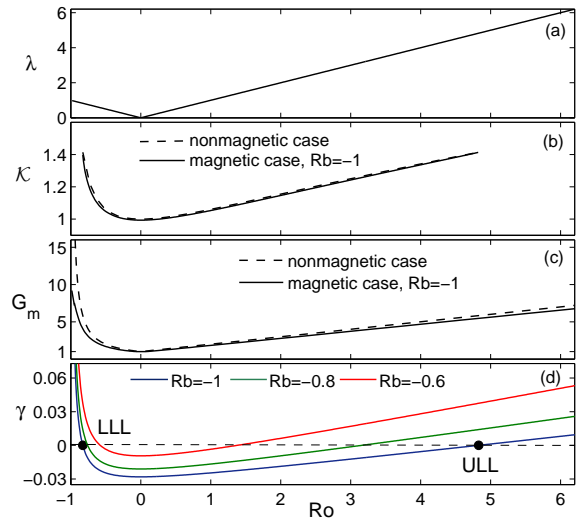}
\caption{(a) numerical abscissa, $\lambda$, (b) Kreiss constant,
${\cal K}$ (c) $G_m$ for HMRI at ${\rm Rb}=-1$ as well as in the
nonmagnetic case and (d) modal growth rate of HMRI from Eq. (6) vs.
${\rm Ro}$ at different ${\rm Rb}=-1, -0.8, -0.6$ and $\alpha=1$.}
\end{figure}

The other relevant notions used to characterize the nonmodal growth
and its connection with the results of modal analysis are the
pseudospectra and numerical range of the nonnormal operator ${\bf
A}$ \cite{Schmid_Henningson2001,Trefethen2005,Schmid2007}. The
maximal protrusion of the numerical range into the upper (unstable)
half in the complex $\omega$-plane -- a numerical abscissa,
$\lambda$, defines the maximum growth rate at the beginning of
evolution (at $t=0^{+}$), $2\lambda=\max_{{\boldsymbol
\psi}(0)}E(t)^{-1}dE(t)/dt|_{t=0^{+}}$. On the other hand, the
extent to which the pseudospectra contours penetrate into the upper
half of the $\omega$-plane determines the amount of transient
amplification over time. This is quantified by the Kreiss constant
${\cal K}=\max_{{\rm Im}(\omega)>0}{\rm Im (\omega)}||({\bf A}+{\rm
i}\omega{\bf I})^{-1}||$, where ${\bf I}$ is the unit matrix and
$||\cdot||$ denotes a suitably defined norm
\cite{Schmid_Henningson2001,Trefethen2005}. This constant provides a
lower estimate for the maximum nonmodal amplification of energy over
time, i.e., $\max_{t>0} G(t)\geq {\cal K}^2$
\cite{Schmid_Henningson2001,Schmid2007}.

Figure 3 shows the normal mode spectra of Eq. (5) and the associated
pseudospectra in the $\omega$-plane at ${\rm Ro}_{\rm Kep}=-0.75$,
where all the eigenfrequencies (thick black dots) are in the lower
half plane, indicating modal stability against HMRI. The mode which
is closer to the ${\rm Im}(\omega)=0$-axis will first cross it and
exhibit HMRI as ${\rm Ro}$ changes beyond the Liu limits, while the
other two modes far in the lower half plane are rapidly damped
magnetic (SMRI) modes. On the other hand, the numerical abscissa and
the frequency, $\omega_K$, that results in the Kreiss constant, lie
in the upper plane, which indicates the nonmodal amplification
larger than ${\cal K}^2$ takes place over intermediate times.

Figure 4, which illustrates the central result of this Letter, shows
(a) the numerical abscissa $\lambda$, (b) the Kreiss constant ${\cal
K}$, (c) the maximum growth $G_m$ for ${\rm Rb}=-1$ and in the
nonmagnetic case as well as (d) the modal growth rate $\gamma$ given
by Eq. (6) at ${\rm Rb}=-1, -0.8, -0.6$ versus ${\rm Ro}$. The
numerical abscissa, measuring the initial optimal growth rate of the
energy, is equal to $|{\rm Ro}|$, i.e., to the maximum growth rate
of ideal SMRI (see also Ref. \cite{Squire2014}) despite the very
high resistivity of the flow. $G_m$ increases linearly with ${\rm
Ro}$ at ${\rm Ro}>0$ and much steeper at ${\rm Ro}<0$ which can be
well approximated by $\propto(1+{\rm Ro})^{-0.78}$. For comparison,
in this plot we also show the maximum transient growth factor for
axisymmetric perturbations in the nonmagnetic case,
$G_m^{(h)}=(1+{\rm Ro})^{{\rm sgn}({\rm Ro})}$, as derived  in
\cite{Afshordi2005}. So, although $G_m$ in the magnetic case is
slightly smaller than that in the nonmagnetic one, the two curves
are in fact close to each other and display nearly the same behavior
with ${\rm Ro}$, a feature that is also shared by the Kreiss
constant (b). Note that the dependencies of $G_m$, $G_m^{(h)}$ (Fig.
4c) and of the modal growth rate $\gamma$ (Fig. 4d) on ${\rm Ro}$
have very similar shapes. Remarkably, the latter, being given by Eq.
(6), can be expressed in terms of the hydrodynamic nonmodal growth
$G_m^{(h)}=(1+{\rm Ro})^{{\rm sgn}({\rm Ro})}$ in the closed form
(${\rm Rb}=-1$)
\begin{equation}
\gamma=\frac{{\rm Ha}^2}{\rm Re}\left[\frac{(G_m^{(h)}+1)^2}
{8 G_m^{(h)}}-1 \right]
\label{connection}
\end{equation}
which is indeed proportional to $G_m^{(h)}$ for larger values. Both
Liu limits, at which HMRI sets in, are therefore connected with a
corresponding threshold  $G_m^{(h)}({\rm Ro}_{\rm
LLL})=G_m^{(h)}({\rm Ro}_{\rm ULL})=5.828$.

In this Letter, we have investigated the nonmodal dynamics of HMRI
due to the nonnormality of a magnetized shear flow with large
resistivity. We traced the entire time evolution of the optimal
growth of the perturbation energy and demonstrated how the nonmodal
growth stage smoothly carries over to the modal behavior at large
times. At small and intermediate (orbital/dynamical) times, HMRI
undergoes transient amplification with the initial growth rate being
equal to that of the most unstable SMRI. Then, it reaches a maximum,
which is higher for larger $|{\rm Ro}|$, and finally at asymptotic
times, it decays or increases exponentially, respectively, when
${\rm Ro}$ lies within or beyond the Liu limits. The transient
growth of HMRI is generally several times larger than its modal
growth during the dynamical time. It also occurs in the Keplerian
regime, where the modal HMRI is thought to be non-existing. As
illustrated in Fig. 4, and quantified exactly in Eq. (7), the modal
growth rate of HMRI displays quite a similar dependence on ${\rm
Ro}$ as the maximum nonmodal growth in the purely hydrodynamic shear
flow, which indicates a fundamental connection between nonmodal
dynamics and dissipation-induced modal instabilities, such as HMRI.
Both, despite the latter being magnetically triggered, rely on
hydrodynamic means of amplification, i.e., they extract energy from
the background flow mainly by Reynolds stress due to shear
\cite{Priede2007}.

This work was supported by the Alexander von Humboldt Foundation and
the German Helmholtz Association in frame of the Helmholtz Alliance
LIMTECH.

\end{document}